\documentclass[%
 amssymb, amsmath,%
 aps,prl,%
 superscriptaddress,
final,%
twocolumn
]{revtex4-1}

\usepackage{bm}%
\usepackage[colorlinks=true]{hyperref}%
\hypersetup{
     colorlinks = true,
     linkcolor = blue,
     anchorcolor = blue,
     citecolor = blue,
     filecolor = blue,
     urlcolor = blue
     }
\usepackage{graphicx}
\expandafter\ifx\csname package@font\endcsname\relax\else
 \expandafter\expandafter
 \expandafter\usepackage
 \expandafter\expandafter
 \expandafter{\csname package@font\endcsname}%
\fi
\usepackage{soul}
\soulregister\SI7
\soulregister\cite7
\soulregister\citeme7
\soulregister\ref7
\soulregister\eqref7
\renewcommand{\hl}[1]{#1}
\renewcommand{\st}[1]{\iffalse{#1}\fi}

\usepackage{xcolor}

\newcommand{\citeme}{$^{\text{\textcolor{red}{??}}}$ }

\newcommand{\IBMANT}{\affiliation{IBM Research at Albany NanoTech, 257 Fuller Road, Albany, New York 12203, USA}}
\newcommand{\SNPSCA}{\affiliation{{Synopsys Inc., 690 East Middlefield Road, Mountain View, California 94043, USA}}}
\newcommand{\SNPSTX}{\affiliation{{Synopsys Inc., 1301 South Mopac Expressway, Bldg. 4, Suite 200, Austin, TX 78746, USA}}}
\newcommand{\SNPSDenmark}{\affiliation{{Synopsys Denmark, Fruebjergvej 3, Postbox 4, DK-2100 Copenhagen, Denmark}}}

\begin{document}

\title{First principles evaluation of fcc ruthenium for use in advanced interconnects}%

\author{Timothy M. Philip}\email{tim@ibm.com}\IBMANT
\author{Nicholas A. Lanzillo}\IBMANT
\author{Tue Gunst}\SNPSDenmark
\author{Troels Markussen}\SNPSDenmark
\author{Jonathan Cobb}\SNPSTX
\author{Shela Aboud}\SNPSCA
\author{Robert R. Robison}\IBMANT

\begin{abstract}
	As the semiconductor industry turns to alternate conductors to replace Cu for future interconnect nodes, much attention has been focused on evaluating the electrical performance of Ru. The typical hexagonal close-packed (hcp) phase has been extensively studied, but relatively little attention has been paid to the face-centered cubic (fcc) phase, which has been shown to nucleate in confined structures and may be present in tight-pitch interconnects. Using \emph{ab initio} techniques, we benchmark the performance of fcc Ru. We find that the phonon-limited bulk resistivity of the fcc Ru is less than half of that of hcp Ru, a feature we trace back to the stronger electron-phonon coupling elements in hcp Ru that are geometrically inherited from the modified Fermi surface shape of the fcc crystal. Despite this benefit of the fcc phase, high grain boundary scattering results in increased resistivity compared to Cu-based interconnects with similar average grain size. We find, however, that the line resistance of fcc Ru is lower than that of Cu below 21 nm line width due to the conductor volume lost to adhesion and wetting liners. In addition to studying bulk transport properties, we evaluate the performance of adhesion liners for fcc Ru. We find that it is energetically more favorable for fcc Ru to bind directly to silicon dioxide than through conventional adhesion liners such as TaN and TiN. In the case that a thin liner is necessary for the Ru deposition technique, we find that the vertical resistance penalty of a liner for fcc Ru can be up to eight times lower than that calculated for conventional liners used for Cu interconnects. Our calculations, therefore, suggest that the formation of the fcc phase of Ru may be a beneficial for advanced, low-resistance interconnects.
\end{abstract}

\date{\today}%
% \revised{December 2014}%
%a feature we trace back to the stronger coupling elements that is geometrically inherited from a less rapidly varying Fermi surface.

\maketitle

% \tableofcontents

\section{Introduction}

% \TMPcomment{Test}
% \NALcomment{Test}
% \TMcomment{Test}
% \TGcomment{Test}
% \RRRcomment{Test}
% \SAcomment{Test}
% \JCcomment{Test}

The introduction of dual damascene Cu wiring in the back-end-of-line (BEOL) provided significant resistance reduction compared to aluminum and enabled the aggressive interconnect pitch scaling witnessed over the last 20 years~\cite{Edelstein2017}. With continued pitch reduction, however, the use of Cu may not be sustainable not only due to the size effect~\cite{Fuchs1938,Sondheimer1952,Mayadas1970,Josell2009}, whereby metal resistivity dramatically increases as cross-sectional area decreases, but also because of electromigration concerns and the reduction of via and line volume occupied by the diffusion and wetting liners necessary for integration of Cu in the BEOL~\cite{Huang2018}. To overcome these challenges posed by the continued use of Cu, research has focused on identifying and evaluating alternative conductors for use in advanced interconnect nodes~\cite{Naeemi2007,Gall2016,Philip2017,VanderVeen2018,Croes2018}. Among the many identified options, Ru has emerged as a promising candidate due not only to reduced impact of the size effect compared to Cu~\cite{Milosevic2018,Dutta2018} but also because of its resilience to electromigration~\cite{Hu2017,Croes2018}.

To understand the performance of Ru-based interconnects, the hexagonal close-packed (hcp) phase of Ru, the dominant phase in bulk Ru, has been extensively studied both experimentally and theoretically using first-principles methods~\cite{Gall2016,Wen2016a,Zhang2016a,Hu2017,Dutta2017a,Dutta2018,Milosevic2018,Dixit2018}. Significantly less attention, however, has been paid to the face-centered cubic (fcc) phase, which has been observed in confined structures such as in striations in tight-pitched interconnect lines and in nanoparticles~\cite{Kusada2013,Hu2017}. As interconnect widths shrink to the tens of nanometers and below, it is increasingly important to understand this less common phase. Studying the properties of fcc Ru can additionally provide valuable insight and direct comparison to other notable cubic BEOL conductors such as Cu, Co, and W~\cite{Cesar2014,Hegde2016,Lanzillo2017a,Lanzillo2018b,Lanzillo2019b}.

In this work, we evaluate the use of fcc Ru for use in the BEOL using \emph{ab initio} techniques. We begin by studying the bulk transport properties of the fcc phase by calculating the phonon-limited resistivity and show that fcc Ru is over twice as conductive as hcp Ru but still 30\% less conductive than Cu. To better understand the performance of this phase in interconnects, we then calculate the reflection coefficients for pertinent grain boundaries using the non-equilibrium Green function method (NEGF). For each grain boundary studied, we find that fcc Ru suffers from a larger reflection coefficient than Cu. We use this information to calculate the expected resistivity of both fcc Ru and Cu over a wide range of grain sizes using the Mayadas-Shatzkes model and see that grain boundary scattering indeed contributes significantly to fcc Ru resistivity for small grain sizes. \hl{We show, however, that fcc Ru offers lower resistivity than other alternate conductors such as hcp Ru, Co, and W.} By accounting for the conductor volume lost to adhesion and wetting liners, \st{however,} we find that the \hl{final }line resistance of fcc Ru interconnects are lower than those made of Cu below line widths of 21 nm. Next, we study the properties of liners that may be necessary for integration of fcc Ru into the BEOL. We benchmark the adhesion performance of the liners by calculating the binding energy of each layer to oxide and to fcc Ru and find that tantalum-based liners have lower binding energies to both oxide and Ru. Notably, we find that the binding energy of Ru directly to silicon dioxide is lower than that of any of the liners studied, which indicates that this phase may adhere to an interlayer dielectric without an adhesion liner
% \NALcomment{The astute reviewer will ask whether or not this same conclusion holds for HCP Ru. }. 
 In the case that a liner is necessary in practice, we calculate the resistance penalty of electrical conduction perpendicular to these liners as would be encountered in BEOL vias. When compared to Cu, we find that the vertical via resistance is dramatically reduced for Ru since the noble metal does not need the wetting layer that Cu requires to achieve void-free gap fill and mitigate electromigration.

\section{Computational Details}
First-principles calculations are performed using the Synopsys QuantumATK software using a double-zeta polarized linear combination of atomic orbitals (LCAO) basis set within the Perdew-Burke-Enzerhof (PBE) generalized gradient approximation (GGA) for the exchange-correlation functional~\cite{smidstrup_quantumatk_2019,QuantumATK2019,Perdew1996a}. All simulated structures are relaxed until the forces on each atom are less than 5 meV/\AA. All reported results are converged with respect to $\mathbf k$-point sampling, and a cutoff energy of 120 Ha is used.

\subsection{Device simulations}
Transport properties are calculated using the non-equilibrium Green function (NEGF) method whereby the transmission function $T(E)$ is given as 
\begin{equation}
	T(E) = \text{Tr}\!\left[G^r(E)\Gamma^L(E) G^a(E) \Gamma^R(E)\right],
\end{equation}
where $E$ is the energy of interest, $G^r(E)$ is the retarded Green function of  describing the system, $G^a(E) = [G^r(E)]^\dag$, \hl{$\Gamma^{L(R)} = i\{\Sigma^{L(R)} - [\Sigma^{L(R)}]^\dag\}$} is the broadening function for the left (right) electrode, and $\Sigma^{L(R)}$ is the contact self-energy for the left (right) electrode~\cite{Brandbyge2002a,Stradi2016a,Hirsbrunner2019}. The retarded Green function is given as 
\begin{equation}
	G^r(E) = \left[ (E+i\eta)S - H - \Sigma^R(E) - \Sigma^L(E)\right]^{-1},
\end{equation}
where $\eta$ is an infinitesimal positive number, $S$ is the overlap matrix for the basis, and $H$ is the Hamiltonian for the scattering region under study. In each structure where transport is calculated, the electrodes are considered to be semi-infinite extensions of the bulk material being studied. To achieve high accuracy in transport calculations, a dense $\mathbf k$-point sampling of 201 by 201 is used in the directions perpendicular to the transport direction. The low-bias, zero temperature conductance $G$ is then directly calculated from the transmission function evaluated at the Fermi energy $E_F$:
\begin{equation}
	G = \frac{2e^2}{h} T(E_F).
\end{equation}
The conductance of the structure is proportional to the cross-sectional area of the device, so we report the area-normalized specific resistivity $\gamma = A/G$ as is standard practice in the literature~\cite{Cesar2014,Lanzillo2017a,Lanzillo2018b,Zhou2018a,Dixit2018,Lanzillo2019}.

\subsection{Bulk transport simulations}
We calculate the bulk phonon-limited resistivity, $\rho$, of fcc Ru directly by solving the semiclassical Boltzmann transport equation~\cite{gunst_first-principles_2016}. The resistivity tensor is given by
\begin{eqnarray}
\frac{1}{\rho_{\alpha \beta}} &=& \int dE\, \sigma_{\alpha \beta}(E) \left(\frac{df^0}{dE}\right)\,
\label{eqn:LinearResponseResistivity}
\end{eqnarray}
where $f^0$ is the Fermi-Dirac distribution function and Cartesian components are labeled by $\alpha, \beta$. The conductivity spectrum $\sigma_{\alpha\beta}(E)$ in Eq.~\eqref{eqn:LinearResponseResistivity} is given by 
\begin{eqnarray}
\sigma_{\alpha \beta}(E) &=& \sum_{n\mathbf{k}} \sigma_{\alpha \beta}(n\mathbf{k}) \delta(E_{n\mathbf{k}}-E)\,,
\end{eqnarray}
and the band conductivity $\sigma_{\alpha \beta}(n\mathbf{k})$ is given as
\begin{eqnarray}
\sigma_{\alpha \beta}(n\mathbf{k}) &=& e^2\tau_{n\mathbf{k}}\mathbf{v}_{\alpha}(n\mathbf{k})\mathbf{v}_{\beta}(n\mathbf{k})\,,
\label{eqn:ConductivitySpectrum}
\end{eqnarray}
where $e$ is the electron charge, $\tau(n\mathbf k)$ is the transport relaxation time for electrons with wavevector $\mathbf k$ in band $n$, $\mathbf v_{\alpha}(n\mathbf k)$ is the electron velocity.

\section{Bulk Characteristics}
We begin by calculating the formation energy of bulk fcc Ru compared to bulk hcp Ru and find that the hcp phase is energetically favorable by just 0.1 eV/atom. This small difference in formation energy explains why the fcc phase can form in confined structures where increased surface energy and strain may overcome this energy barrier. Additionally, this result suggests that the fcc phase may be prominent in advanced interconnect nodes where line widths will be less than 10 nanometers and the standard BEOL thermal budget of 300-400$^\circ$C could provide enough energy for the fcc phase to stabilize. It is therefore paramount to characterize the electrical properties of this phase to benchmark how it compares to that of Cu and other alternate conductors. 

\begin{figure}[t!]
	\includegraphics[width=\columnwidth]{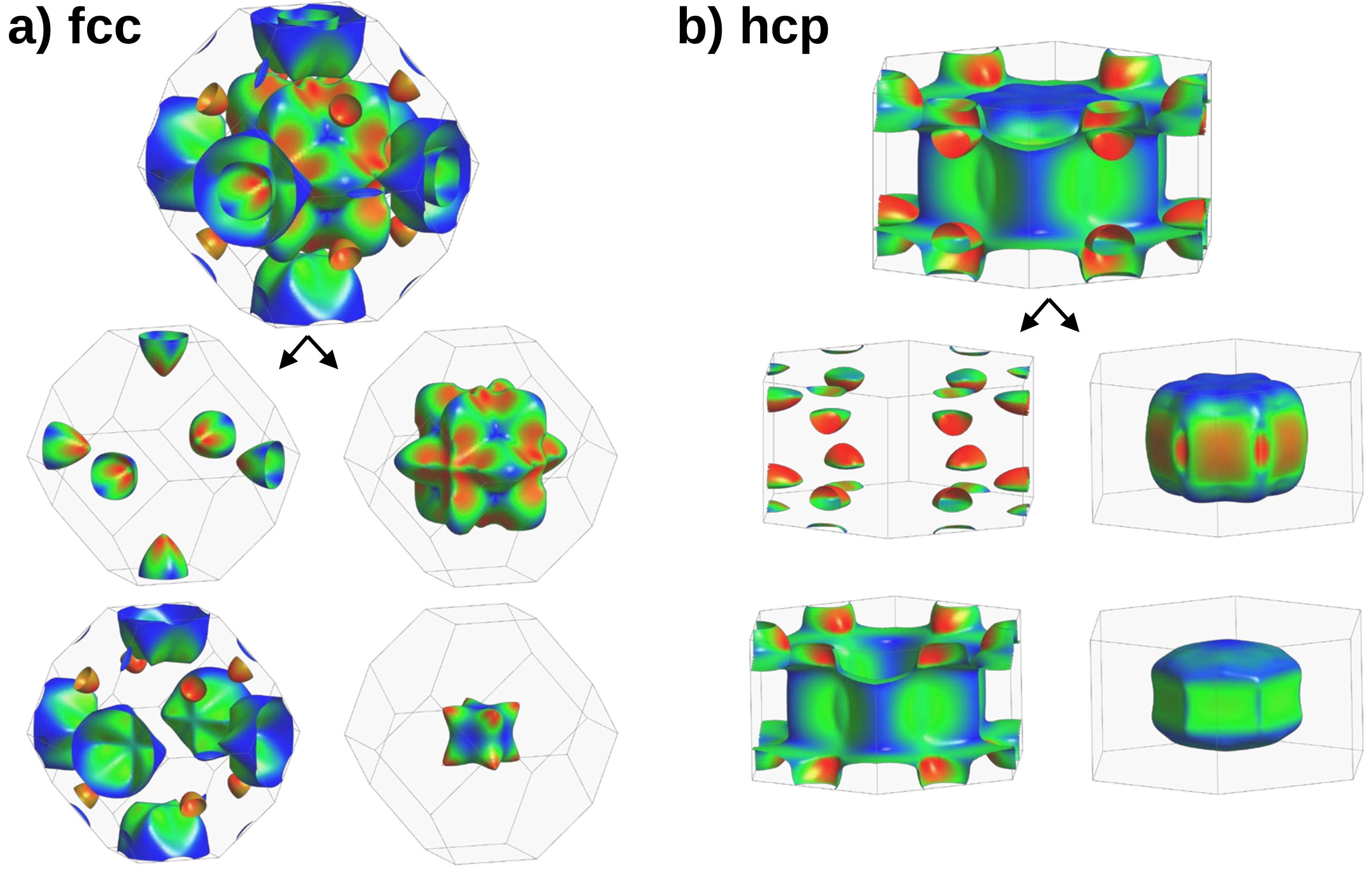}
	\includegraphics[width=\columnwidth]{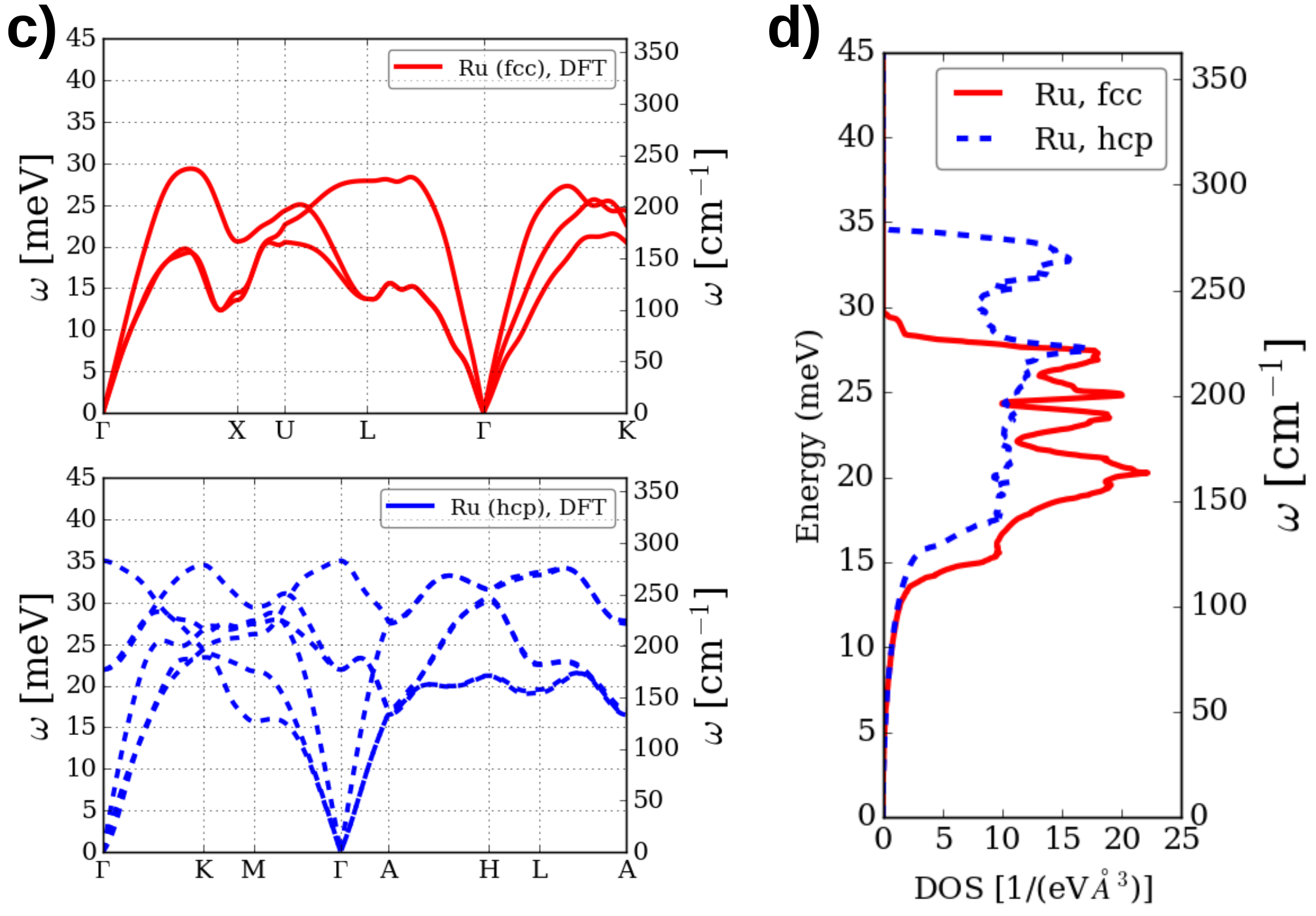}
	\caption{Properties of bulk Ru; Total Fermi surface and contributions from the four conduction bands for fcc (a) and hcp (b) phases. Colors illustrate the gradient (the spatially resolved Fermi velocity). The phonon dispersions are shown in (c) and phonon density of states in (d).}\label{fig:bulkRu}
\end{figure}

To this end, we calculate bulk transport properties of the fcc and hcp phases. For most alternate conductors, the bulk resistivity is typically an experimental value that can be used as an input to various resistivity models to understand performance in interconnects. \hl{Such measurements, however, have not yet been performed for fcc Ru since it is unstable in bulk films, so we instead solve the linearized Boltzmann transport equation to obtain the bulk resistivity.}\st{Since the fcc phase is unstable in bulk films, acquiring this value experimentally is challenging. Instead, we solve the linearized Boltzmann transport equation.}
%%%%%%%%%%%%%%%%%%%%%%%%%%%%%%%%%%%%%%%%%%%%%%%%%%%%%%%%%%%%%%%%%%%%%%%%%
To perform the simulation we employ a two\hl{-}step procedure~\cite{smidstrup_quantumatk_2019}, where we first perform a separate calculation of the scattering rate for a selection of  approximately $100\,\mathbf{k}$-points within $2\,\textrm{meV}$ from the Fermi surface. This is integrated to give the relaxation time at a specific energy:
\begin{eqnarray}
\frac{1}{\tau(E)} = \frac{1}{n(E)}\sum_{n\mathbf{k}}\frac{1}{\tau_{n\mathbf{k}}} \delta(E_{n\mathbf{k}}-E) ,
\label{eqn:ScatteringRateEnergyDependent}
\end{eqnarray}
where $n(E)$ is the density of states. 
We then calculate the constant relaxation time result by assuming $\tau(n\mathbf k) \to \tau$ when solving Eq.~\ref{eqn:LinearResponseResistivity} from a fine $\mathbf{k}$-point integration on a $22\times22\times22(11)$ mesh for the fcc (hcp) phase.
Importantly, the product $\tau \rho_{\alpha \beta}$ is a Fermi surface property independent of the scattering mechanisms. We can therefore separate resistivity changes as coming from either the Fermi velocity and surface variation or from the specific phonon-limited scattering rate.

The key properties of bulk Ru are collected in Fig.~\ref{fig:bulkRu}. Figure~\ref{fig:bulkRu}(a, b) illustrates the Fermi surfaces and the four bands contributing in fcc and hcp Ru, respectively. Compared to the hcp phase, we see that the surfaces for the fcc bands exhibit more fluctuations in $\mathbf{k}$-space. For both phases, we calculate the $\rho \tau$ product directly from tetrahedron integrations of these surfaces, thereby accounting for the full anisotropy of the Fermi velocity on the surface depicted. In Fig.~\ref{fig:bulkRu}(c, d) we show the phonon dispersion as a function of phonon wavevector $\mathbf{q}$ and phonon density of states, which are calculated using s $9 \times 9 \times 9$ super cell. The results for the hcp phase compares well with previous simulations and experiments~\cite{palumbo_lattice_2017, heid_anomalous_2000}.

To obtain an approximation for scattering time in the fcc and hcp phases we calculate from first-principles the electron-phonon coupling and derived scattering rates to  solve the linearized Boltzmann transport equation fully numerically within the full-band relaxation time approximation. Details of the methodology are described in Ref.~\onlinecite{gunst_first-principles_2016}. 
Such DFT calculation of the resistivity of metals is computationally demanding 
as one needs to integrate the electron-phonon coupling over both electron and phonon wave vectors ($\mathbf{k}$- and $\mathbf{q}$-space, respectively),
and only few studies of the electron-phonon coupling in metals exist that includes a full integration~\cite{bauer_electron-phonon_1998,Gall2016,markussen_electron-phonon_2017,smidstrup_quantumatk_2019}.
For the integration of the scattering rate we use a sampling of $30 \times 30 \times 30$ $\mathbf{q}$-points and tetrahedron integration.
We then integrate the scattering time according to Eq.~\ref{eqn:ScatteringRateEnergyDependent} to obtain a constant scattering time at the Fermi level for both the hcp and fcc phases.

\begin{table}[t!] 
	\begin{tabular*}{\columnwidth}{l @{\extracolsep{\stretch{1}}}*{6}{c}@{}}
		\hline\hline
		Metal & $\rho \tau$ ($10^{-12}\,\Omega$-m-s) &$\tau$ (fs) & $\lambda$ [nm] & $\rho$ ($\mu\Omega$-cm) \\ \hline
		Ru (fcc) & 6.8 & 17.9 & 6.5 & 3.8 \\
		Ru (hcp) & 10.5 (7.7) & 9.4 & 5.8 & 11.1 (8.2) \\
		Cu (fcc) & 6.04\cite{Gall2016} & 36\cite{Gall2016} & 39.9\cite{Gall2016} & 1.68\cite{Gall2016} \\\hline\hline
	\end{tabular*}
	\caption{Bulk resistivity and scattering time for fcc Ru, hcp Ru, and fcc Cu. For hcp Ru, the value in parentheses is calculated parallel to the hcp $c$-axis while those not in parentheses is in the direction perpendicular to the $c$-axis. The experimental reference value for hcp-Ru is $7.8\,\mu\Omega$-cm~\cite{Gall2016}.}\label{tab:bulk_rho}
\end{table}

Table~\ref{tab:bulk_rho} summarizes the $\rho\tau$ product, phonon-derived scattering time $\tau$, mean free path $\lambda$ and extracted bulk resistivity $\rho$.
For the hcp phase we find a resistivity value 8.2-11.1$\,\mu\Omega$-cm which is 5-40\% greater than the experimental reference value of $7.8\,\mu\Omega$-cm. This is fairly good agreement considering the full first-principles approach utilized, and we will therefore use the simulations to compare the performance of the two phases.
We find that the fcc phase of Ru has a $\rho\tau$ product 13-35\% lower than that of hcp Ru, which combined with the longer scattering time results in a bulk resistivity that is 2-3 times lower indicating that fcc phase of Ru is a superior conductor to the typical hcp phase. In comparison to Cu, we find that the $\rho\tau$ product of fcc Ru is about equal, but the shorter scattering time results in fcc Ru having twice as high a bulk resistivity. 

While the $\rho\tau$ product does provide the correct trend, it is insufficient to account for the full discrepancy between the two phases of Ru. 
To explain the lower scattering rate of the fcc phase we explore the elastic model of acoustic phonon scattering in metals.
%The transition rate is given by
%\begin{eqnarray}
%P^{\lambda n n'}_{\mathbf{k}\mathbf{k}' \mathbf{q}}&=&\frac{2\pi}{\hbar} |g_{\mathbf{k}\mathbf{k}'\mathbf{q}}^{\lambda n n'}|^2
%n_{ \lambda \mathbf{ q} }\, \delta \! \left(\varepsilon_{n'\mathbf{k}'}-\varepsilon_{n\mathbf{k}}-\hbar \omega_{\lambda\mathbf{q} } \right), \label{eqn:FGRtransitionRate} \\
%\bar{P}^{\lambda n n'}_{\mathbf{k}\mathbf{k}' \mathbf{q}}&=&\frac{2\pi}{\hbar}  |g_{\mathbf{k}\mathbf{k}'-\mathbf{q}}^{\lambda n n'}|^2
%(n_{\lambda -\mathbf{q}}+1) \delta \! \left(\varepsilon_{n'\mathbf{k}'}-\varepsilon_{n\mathbf{k}}+\hbar \omega_{\lambda-\mathbf{q} } \right), \nonumber
%\end{eqnarray}
%where $n_{\lambda \mathbf{q} }$ is the phonon occupation operator,
%and $g_{\mathbf{k}\mathbf{k}'\mathbf{q}}^{\lambda n n'}$ the EPC constant from \eqref{eq:M-definition}.
Within the Ziman model, where the acoustic coupling constant is expressed by $M(\mathbf{k},\mathbf{q}) = \hbar q D_\text{ac}^2 /(2 \mu v_{ph})$, the scattering rate can be expressed as:\cite{ZimanBook}
\begin{eqnarray}
%\frac{1}{\tau} \approx \frac{3 \zeta(3)}{\pi}\frac{\Delta^2 T^3}{\hbar^4 \mu v_{ph}^4 v_F},
\frac{1}{\tau} \approx \frac{7 \zeta(3)}{8 \pi}\frac{D_\text{ac}^2 (k_B T)^3}{\hbar^4 \mu v_{ph}^4 v_F}.
\label{eqn:ZimanLargeFermi}
\end{eqnarray}
Here $\mu$ is the mass density, \hl{$\zeta$ is the Riemann zeta function}, $v_{ph}$ is the sound velocity, $v_{F}$ is the Fermi velocity, \hl{$D_\text{ac}$ is the acoustic deformation potential,} and we have assumed the limit of a large Fermi surface compared to the thermal phonon momentum $k_F>>k_B T/(\hbar v_{ph})$.
The main factors that could modify the scattering rate are the effective velocity of sound and the geometrical effect from the Fermi surface shape that could modify the deformation potential.

From the phonon dispersion and density of states in Fig.~\ref{fig:bulkRu}(c, d), we see that while some phonon softening occurs in the fcc phase that reduces the effective phonon frequencies and phonon velocity $v_{ph} \approx \omega(q)/q$, it is less important for the acoustic modes at low phonon momentum and hence cannot explain the difference in scattering rate.
The discrepancy therefore traces back to the actual coupling matrix element\st{, $\Delta$,} and possibly the full momentum dependence which was included in the actual simulations.
Due to the more isotropic nature of the hcp Fermi surface, the selection rules for scattering can be more easily fulfilled, and the coupling strength of the hcp phase therefore exceeds that of the fcc phase.
%This clearly illustrates that coupling strengths are highly symmetry dependent and this relation between the phonon coupling strength and the Fermi surface shape of metals could be more generic and favoring complex and sharp-featured Fermi surfaces for reduced coupling.
We therefore can explain the resistivity reduction partly from (i) Fermi velocity and surface variation included in the $\rho \tau$ product and (ii) the dominating geometrical effect from the shape of the Fermi surface enabling stronger coupling elements for the hcp phase.

%%%%%%%%%%%%%%%%%%%%%%%%%%%%%%%%%%%%%%%%%%%%%%%%%%%%%%%%%%%%%%%%%%%%%%%%%%%%%%%%%%%%%%%%%%%%%%%%

\subsection{Grain boundary scattering}

To further compare the bulk transport properties of fcc Ru to Cu, we quantify the impact of grain boundary scattering, which contributes a significant amount to resistance of nanoscale interconnects~\cite{Mayadas1970,Hegde2016}. To this end, we calculate the reflection coefficients of four representative coincident site lattice twin grain boundaries. We use the standard $\Sigma$ notation for fcc twin grain boundaries where $\Sigma = \delta (h^2 + k^2 +\ell^2)$, $(hk\ell)$ corresponds to the Miller indices of the surface that defines the grain boundary,  $\delta = 1$ when the summation is odd, $\delta = 1/2$ when the summation is even. The reflection coefficient $r$ is calculated by computing the specific resistivity of each grain using the NEGF method:
\begin{equation}
	r = 1 - \frac{T_{GB}(E_F)}{T_{\langle hk\ell\rangle}(E_F)} = 1 - \frac{\gamma_{\langle hk\ell\rangle}}{\gamma_{GB}},
\end{equation}
where $T_{GB}(\gamma_{GB})$ is the transmission function (specific resistivity) of the grain boundary and $T_{\langle hk\ell\rangle}(\gamma_{\langle hk\ell\rangle})$ is the ballistic transmission function (specific resistivity) in the direction normal to the grain boundary.

\begin{table}[t!] 
	\begin{tabular*}{\columnwidth}{l @{\extracolsep{\stretch{1}}}*{8}{c}@{}}
		\hline\hline
		      &    \multicolumn{2}{c}{$\Sigma 3$}  &    \multicolumn{2}{c}{$\Sigma 5$} &    \multicolumn{2}{c}{$\Sigma 9$} &    \multicolumn{2}{c}{$\Sigma 11$} \\ \cline{2-3} \cline{4-5} \cline{6-7} \cline{8-9}
		Metal & $\gamma_{GB}$  & $r$ & $\gamma_{GB}$  & $r$ & $\gamma_{GB}$  & $r$ & $\gamma_{GB}$  & $r$   \\ \hline
		Ru (fcc)        & 7.67 & 0.26 & 9.73 & 0.46 & 9.68 & 0.50 & 8.66 & 0.43    \\
		Cu (fcc)\cite{Lanzillo2017a} & 0.22 & 0.02 & 1.32 & 0.13 & 1.80 & 0.14 & 0.64 & 0.07 \\\hline\hline
	\end{tabular*}
	\caption{Values for specific resistivity (in units of $10^{-12}\, \Omega$-cm) and reflection coefficient for representative  coincident site lattice twin grain boundaries in fcc Ru and Cu. %\NALcomment{Depending on how rigorous / comprehensive of a comparison you're looking for, you could also include the same GB reflection coefficients from Tianji's AIP paper (aluminum) and/or my JAP paper (Pt, Pd, Rh, Ir) }
	}\label{tab:refl_coeff}
\end{table}

\begin{figure}[t!]
	\includegraphics[width=0.9\columnwidth]{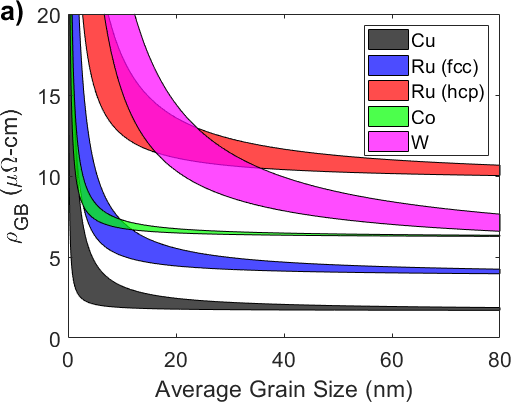}

	\includegraphics[width=0.9\columnwidth]{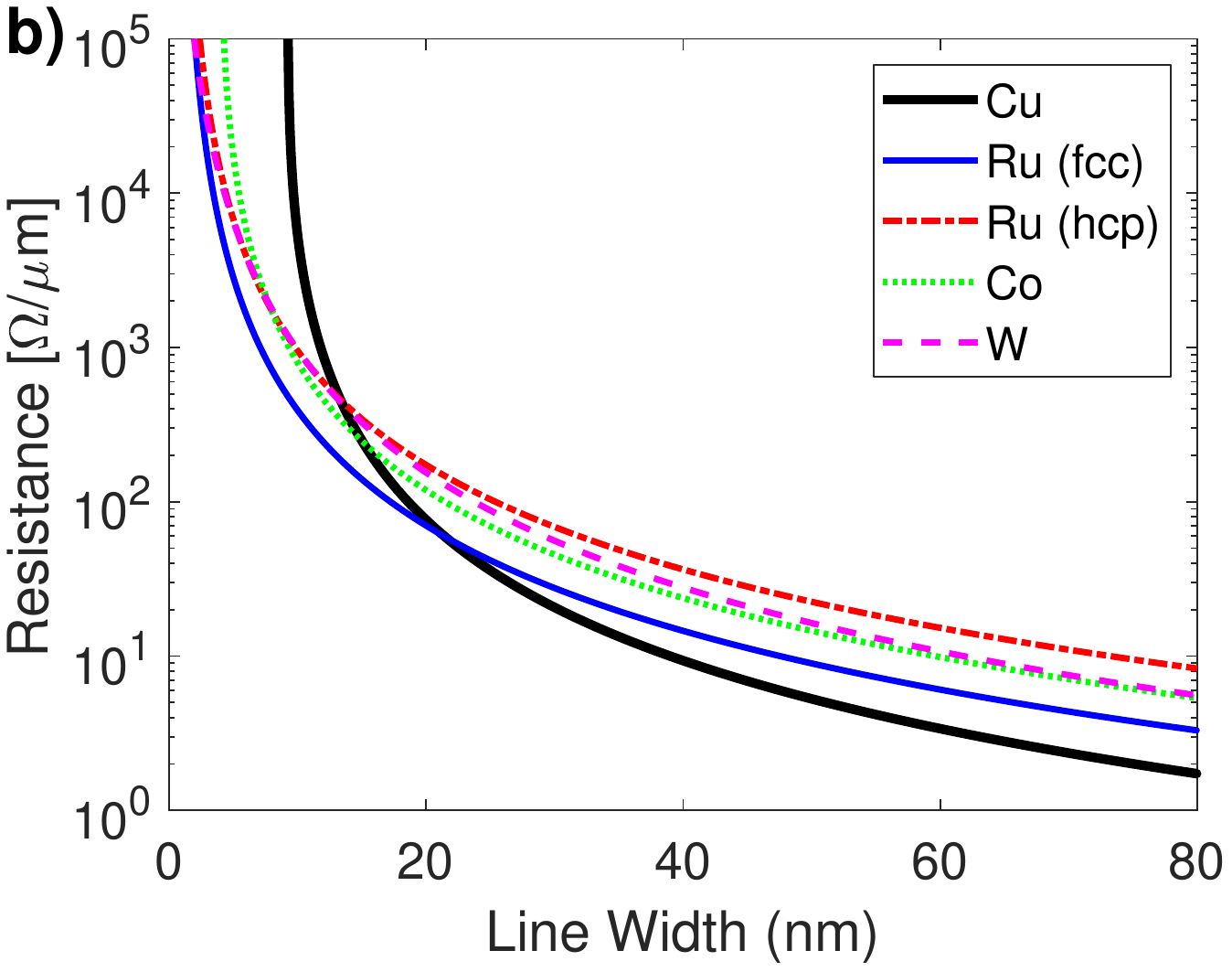}
	\caption{(a) Resitivity of Cu, fcc Ru, \hl{hcp Ru, Co, and W} as a function of average grain boundary using the Mayadas-Shatzkes model. For \st{Cu}\hl{each metal, the} shaded region covers \hl{the range of }reflection coefficients \st{ranging from 0.02 to 0.14 and for Ru, 0.26 to 0.43}\hl{computed in this work and previous studies~}\cite{Lanzillo2017a,Lanzillo2018b,Lanzillo2019b}. (b) Expected line resistance due to grain boundary scattering. \hl{For each metal, we assume state-of-the-art liner and barrier thicknesses to accurately compare the line resistances.}\st{We assume that the Cu lines have a total liner thickness of 3 nm and that the Ru lines have a total liner thickness of 0.3 nm.}}\label{fig:MS_resitivity}
\end{figure}

Table~\ref{tab:refl_coeff} summarizes the computed specific resistivity and reflection coefficient for grain boundaries $\Sigma 3$, $\Sigma 5$, $\Sigma 9$, and $\Sigma 11$. Like many other fcc metals, we see that for both Ru and Cu, the reflection coefficients follow the trend that $r_{\Sigma 3} \ll  r_{\Sigma 11} <r_{\Sigma 5} \approx r_{\Sigma 9} $, which can be understood by examining the symmetry and void structure of the interface of each grain boundary~\cite{Lanzillo2017a}. Notably, for each grain boundary studied here, the reflection coefficient for Ru is much larger than that for Cu indicating that grain boundary scattering plays a much larger role in the resistance of fcc Ru lines. To quantify this impact, we input our computed reflection coefficients into the Mayadas-Shatzkes empirical resistivity model~\cite{Mayadas1970}, which is given as
\begin{equation}
	\rho_{GB} = \frac{\rho_0}{1-3/2\alpha + 3\alpha^2 - 3\alpha^3 \ln (1 + 1/\alpha)},
\end{equation}
where $\rho_{GB}$ is the modified resistivity after accounting for grain boundary scattering, $\alpha = \frac{\lambda}{d} \frac{R}{1-R}$, $\lambda$ is the electron mean free path, $d$ is the average grain size, and $R$ is the average grain boundary reflectivity. Since experimental evaluation of the mean free path of fcc Ru is difficult, we calculate the mean free path using a constant mean free path approximation of Eq.~\eqref{eqn:ScatteringRateEnergyDependent}:
\begin{eqnarray}
\frac{1}{\lambda(E)} = \frac{1}{n(E)}\sum_{n\mathbf{k}}\frac{1}{|\mathbf{v}(n\mathbf{k})| \tau_{n\mathbf{k}}} \delta(E_{n\mathbf{k}}-E) ,
\label{eqn:MFPEnergyDependent}
\end{eqnarray}
where $\tau_{n\mathbf{k}}$ and $\mathbf{v}(n\mathbf{k})$ are the previously calculated band resolved scattering time and electron velocities. The resulting $\lambda(E_F)$ values are given in Table~\ref{tab:bulk_rho}.

Figure~\ref{fig:MS_resitivity}(a) shows the calculated resistivity of \st{both} fcc Ru \hl{against that of Cu, hcp Ru, Co, and W}\st{and Cu} over a wide range of grain sizes\st{ and values of R ranging from 0.26 to 0.43 for Ru and 0.02 to 0.14 for Cu}. \hl{The values for bulk resistivity and mean free path of Cu, hcp Ru, Co, and W are taken from \cite{Gall2016}, and the shaded regions for each metal indicate the range of reflection coefficients calculated here for Ru and from previous studies for the Cu, Co, and W~\cite{Lanzillo2017a,Lanzillo2018b,Lanzillo2019b}.} For large grain sizes where grain boundary scattering is negligible compared to inelastic phonon scattering, we see that the expected resistivity is strongly clustered around the values of bulk conductivity for each metal. As the grain sizes become smaller, the spread of resistivity values increases due to the increasing influence of grain boundary scattering. \hl{When compared to Cu, }we see that because of both the larger bulk resistivity and the increased reflectivity of the grain boundaries studied, fcc Ru continues to have higher resistivity for equal average grain size. \hl{Compared to the other alternate conductors, however, we see that fcc Ru provides much lower resistivity than both hcp Ru and W over the entire range of grain sizes. Only Co has potential to offer  lower grain boundary resistivity compared to fcc Ru but only in the case that the average grain size is smaller than about 5 nm.}
% Although grain size typically decreases with film thickness indicating that Cu should remain less resistive than fcc Ru for narrower interconnects~\cite{Steinhogl2005,Lanzillo2018b}, recent work based on subtractive Ru interconnects suggests that larger grain sizes can be achieved with Ru than can normally be attained with Cu~\cite{Wan2018,Yoon2019}.

%Since grain size has shown to decrease with film thickness,\cite{Steinhogl2005,Lanzillo2018b} we expect a similar trend in occur with interconnect width. 
The actual resistance of \st{fcc Ru lines compared to Cu lines}\hl{interconnects}, however, \st{will }depends not only the average grain size for each metal at the dimension of interest but also on the volume of the conductor that is occupied by adhesion and wetting layers~\cite{Clarke2014}. In Fig.~\ref{fig:MS_resitivity}(b), we calculate the expected line resistance for \st{Cu and Ru }interconnects \hl{ of the same set of metals compared in Fig.~\ref{fig:MS_resitivity}(a)} assuming that the average grain size is equal to the line width, the reflection coefficient is an average of those calculated for each metal\st{ in Table~\ref{tab:refl_coeff}}, and the aspect ratio of the lines is 2.0. \hl{For each metal, we assume the thinnest liner reported in the literature to represent the most optimistic line resistance. } For Cu lines, we assume a liner of 3 nm for the bottom of the trench and 2.3 nm for each sidewall,~\cite{Lanzillo2019a} while for Ru lines, we assume a 0.3 nm liner on each side~\cite{Wen2016a}. \hl{A 1 nm liner is assumed for Co lines, since reliability failure has been reported for thinner liners.\cite{Bekiaris2017} Tungsten has shown the potential for linerless deposition, so we assume no liner in this calculation.\cite{Bakke2016}}
We note that the line resistances presented here exclude the effect of surface scattering, which is highly dependent on the surface microstructure of fabricated lines, and therefore nominal resistance values reported here are optimistic~\cite{Fuchs1938,Sondheimer1952}. Nevertheless, if surface scattering is similar between Cu and Ru lines, the relative trends should remain consistent. 
For large line widths, we see that Cu offers lower line resistance than \st{equivalent width fcc Ru interconnects}\hl{than the alternate conductors} as is expected due to its lower bulk resistivity. 
\hl{Below about 21 nm, we see that the alternate conductors begin to offer lower line resistance than Cu as the adhesion and wetting layers necessary for Cu wiring consume a non-negligible percentage of the wire cross-sectional area. Although the lower line resistance of fcc Ru compared to hcp Ru and W is expected based on the lower line resistivity exhibited in Fig.~\ref{fig:MS_resitivity}(a), we find that fcc Ru demonstrates lower line resistance than Co for all line widths studied. The thicker liner needed for Co reliability increases the effective line resistance of Co interconnects and results in the clear separation in the line resistance of Co and fcc Ru shown in Fig.~\ref{fig:MS_resitivity}(b). }
\st{Below about 21 nm line width, however, we find that fcc Ru is less resistive than copper. At such confined dimensions, the adhesion and wetting layers necessary for Cu wiring consume a non-negligible percentage of the wire cross-sectional area and results in smaller grains compared Ru lines of the same line width.} Our calculations therefore indicate that below 21 nm line width, fcc Ru interconnects may provide significant resistance reduction compared to equivalent width Cu \hl{and alternate conductor} lines.      % \NALcomment{There are some references out there demonstrating how large grains can be grown in Ru interconnects. You could also find some papers (including Liyang's MRS from a few years ago) showing that grain in Cu are fairly small. With this, you could make the argument (even indicating on the graph of Fig1) even though Ru has higher reflection coefficients, the resistivity could still be lower than Cu. }

\begin{figure}[t!]
	\includegraphics[width=\columnwidth,clip=true,trim={0.75in 3.5in 0.75in 3.5in}]{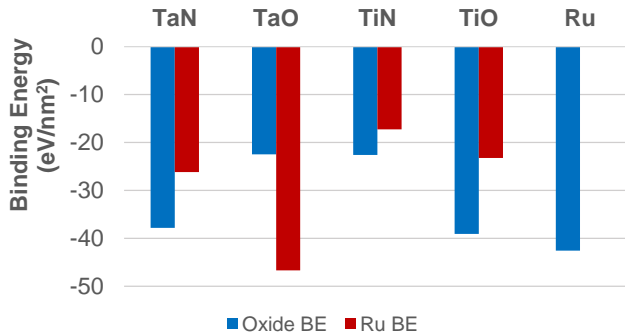}
	\caption{Binding energies of various adhesion liners to both silicon dioxide (Oxide BE) and fcc Ru (Ru BE). For comparison, the binding energy of fcc Ru directly to silicon dioxide is also include}\label{fig:binding_energy}
\end{figure}

\section{Adhesion and Electrical Properties of Liners}

%Next we study the properties of adhesion liners that may be required for the integration of fcc Ru into the BEOL. 
For the integration of Cu into the modern damascene process, a thin diffusion liner and wetting liner are necessary to prevent the electromigration and promote nucleation of Cu~\cite{Hu1986,Hu1995,Hu1998,Edelstein2017}. Ruthenium is not expected to require a wetting layer due to its extremely low susceptibility to electromigration, but a thin adhesion liner such as TaN or TiN may be necessary to adhere the metal to the dielectric~\cite{Zhang2016a}. In the following we evaluate both the adhesion properties of candidate liner materials to fcc Ru and silicon dioxide and the via resistance penalty associated with their integration.

To quantify the adhesion properties of Ru to dielectric, we calculate the binding energy of various liners to both the alpha phase of SiO$_2$ and Ru. We consider both TaN and TiN along with their oxides TaO and TiO, which represent the worst-case scenario in which the liners are fully oxidized during deposition. In these calculations, we assume that the (100) surface of the oxide or Ru is bonded to the (100) surface  of the liner material. The binding energy is then given by 
\begin{equation}
	E_\text{binding} = E_\text{interface} - E_\text{liner} - E_\text{oxide/Ru},
\end{equation}
where $E_\text{interface}$ is the total energy of the interface between two materials, $E_\text{liner}$ is the energy of the liner surface without the presence of the oxide or Ru, and $E_\text{oxide/Ru}$ is the energy of the oxide or Ru surface without the presence of the liner. All reported binding energy values are normalized by the area of the interface to give a metric that is independent of simulation cell size. 

Figure~\ref{fig:binding_energy} summarizes the binding energy of liners TiN, TaN, TiO, and TaO to both fcc Ru and silicon dioxide.
%\NALcomment{a choice of phrasing: are all of these candidate liners? or do you want to view the TaO and TiO as "worst-case" oxidized versions of TaN and TiO. If that's the case, it could be an additional selling point. Even if TaN and TiO become oxidized, the via resistance is not substantially impacted. They also increase binding strength!}
In addition, we calculate the adhesion of the (100) surface of fcc Ru directly to silicon oxide for comparison. Overall, we see that all four liners have a negative binding energy to both Ru and oxide, which tells us that all structures energetically favorable to form. The tantalum-based liners offer a lower binding energy than the titanium-based ones indicating that TaN or TaO offer better adhesion performance than TiN or TiO. Notably, we see that fcc Ru naturally has a more favorable binding energy to oxide than any of the candidate liners studied here, which suggests that fcc Ru may not require an adhesion liner at all. If indeed the requirement for an adhesion liner can be removed for integration of Ru into the BEOL, significant via resistance reductions may be achieved. %\NALcomment{Another interesting conclusion here is that Ru appears to bond more strongly when N is replaced with O. This seems to be true for both TaN and TiN. Perhaps the Ru-O bond energy is stronger than the Ru-N bond energy? }

\begin{table}[t!] 
	\begin{tabular*}{\columnwidth}{l @{\extracolsep{\stretch{1}}}*{3}{c}@{}}
		\hline\hline
		Structure 				& $T(E_F)$ 	& Area (\AA$^2$)& $\gamma$ ($10^{-12}\, \Omega$-cm$^2$)    \\ \hline
		Ru/TaN/Ru       		&  2.91		& 37.63			&  16.70   \\
		Ru/TaO/Ru       		&  3.18		& 37.63			&  15.25   \\
		Ru/TiN/Ru       		&  2.47		& 37.63			&  19.65   \\
		Ru/TiO/Ru       		&  2.39		& 37.63			&  20.31   \\
		Cu/TaN/Cu\cite{Zhou2018a} 		&  1.06 	& 32.72			&  28.03   \\
		Cu/Ta/Ru/Cu\cite{Lanzillo2019}		&  0.99 	& 26.13			&  34.24   \\
		Cu/TaN/Ru/Cu\cite{Lanzillo2018c}		&  0.28 	& 26.11			&  122.3   \\
		Cu/TaO/Ru/Cu\cite{Lanzillo2019} 	&  0.66 	& 26.13			&  51.29   \\
		\hline\hline
	\end{tabular*}
	\caption{Vertical resistance of various adhesion liners for Ru compared to that of diffusion and wetting layers used for Cu.}\label{tab:via_R}
\end{table}

% \begin{figure}[t!]
% 	{\includegraphics[width=\columnwidth,clip=true,trim={0.2in 0.5in 4.5in 0.0in}]{via_structure.pdf}}
% 	\caption{elaxed atomic structures used for vertical via resistance calculations through (a) TaN, (b) TaO, (c) TiN, and (d) TiO liners. The teal atoms are Ru, the light blue atoms are Ta, the dark blue atoms are N, the red atoms are oxygen, and the white atoms are Ti. }\label{fig:via_structure}
% \end{figure}
\begin{figure*}
	{\includegraphics[width=\textwidth,clip=true,trim={0.0in 4.2in 0.5in 0.0in},page=2]{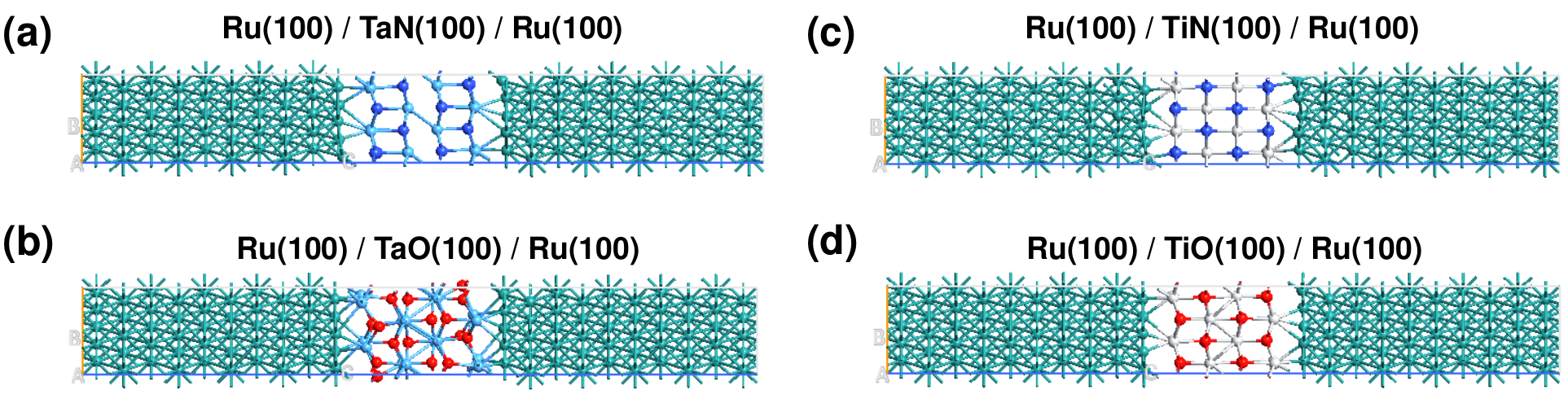}}
	\caption{Relaxed atomic structures used for vertical via resistance calculations through approximately 1 nm of (a) TaN, (b) TaO, (c) TiN, and (d) TiO liners. The teal atoms are Ru, the light blue atoms are Ta, the dark blue atoms are N, the red atoms are oxygen, and the white atoms are Ti. }\label{fig:via_structure}
\end{figure*}

As noted, the presence of an adhesion liner plays a critical role in determining BEOL parasitic RC delay since the resistance of interconnects depends not only on the bulk resistivity calculated in the previous section but also on the ``vertical resistance'' of current flow through vias that connect a given metal level to the one above or below it. For Cu interconnects, an excess of 100 $\Omega$ of vertical via resistance can be expected for future technology nodes due to the presence of diffusion barriers and wetting layers~\cite{Lanzillo2018c,Lanzillo2019}. Although we find that the binding of pure fcc Ru to oxide is more favorable than the liners we studied, a liner may still be required to promote growth of Ru depending on the deposition method and chemistry of the precursors~\cite{Li2007,Zhang2016a,VanderVeen2018}. It is therefore valuable to quantify the via resistance penalty due to the presence of an adhesion liner.

Using the NEGF method, we calculate the vertical resistance of the candidate liners by computing the transmission function of the relaxed heterostructures depicted in Fig.~\ref{fig:via_structure}. In each structure, transport is assumed along the $\langle100\rangle$ direction through each material. A thin barrier of a candidate liner material approximately 1 nm thick is placed in the center of each structure to represent the liner that would be present at the bottom of a traditional damascene via. The electrodes are taken to be semi-infinite extensions of $\langle100\rangle$ fcc Ru, and the transverse directions are considered to be periodic to model bulk-like conduction parallel to the interface. Although we model an idealized, single-crystalline interface that may not \emph{a priori} appear physically accurate, previous studies demonstrate that the resistance calculated using first principles techniques through such structures quantitatively agree with experimental measurements~\cite{Lanzillo2018b,Lanzillo2018c}.

In Table~\ref{tab:via_R}, we summarize the calculated transmission per spin at the Fermi energy, the cross-sectional area of the structure, and specific resistivity of each fcc Ru heterostructure along with reference values from previous study of vertical resistance through Cu diffusion and wetting liners. For the fcc Ru liners, we find that tantalum-based liners provide the lowest via specific resistivity of about $16 \times 10^{-12}\,\Omega$-cm$^2$. Titanium-based liners result in an approximately 25\% higher via resistance of about $20 \times 10^{-12}\,\Omega$-cm$^2$. Although oxidation typically results in a more resistive material, our results show that even if a TiN liner is fully oxidized to TiO, we should only expect an increase of resistance of about 3\%. For the case where TaN is fully oxidized, forming TaO, we find that the via resistance is \emph{lowered} by almost 10\%, which follows the results seen in Cu liner calculations~\cite{Lanzillo2018c,Lanzillo2019}. These results indicate the via resistance of Ru interconnects are less prone to degradation compared to Cu. Overall, our results suggest that tantalum-based liners with their superior adhesion properties and low via resistance offer the best adhesion liner performance for future fcc Ru interconnects.

Comparing our vertical via resistance calculations for fcc Ru to those of previous results studies of Cu diffusion and wetting layers illustrates an important benefit of Ru-based interconnects. Although we find that fcc Ru suffers more from grain boundary scattering than Cu, here we see the opposite behavior for via resistance. Although the one nanometer liners used in these calculations are idealized and highly scaled compared to what is currently used on silicon, a quantitative comparison between similar calculations can be illuminating. The combination of Cu/TaN/Ru/Cu, a typical state-of-the-art BEOL liner and wetting layers stack, results in a vertical resistance of $122.3 \times 10^{-12}\,\Omega$-cm$^2$~\cite{Lanzillo2018c}, a specific resistivity that is over eight times that which we calculate for the TaO adhesion liner for fcc Ru. Even in the optimistic scenario without nitridation or oxidation, the stack of Cu/Ta/Ru/Cu results in a vertical resistance of $34.24 \times 10^{-12}\,\Omega$-cm$^2$~\cite{Lanzillo2019}, which is still twice as resistive as a TaO Ru liner. Much of the vertical resistance seen in Cu vias can be attributed to the fact that both a diffusion barrier and a wetting layer are required for BEOL integration. By removing the Ru wetting layers, the Cu/TaN/Cu specific resistivity is only $28.03\times 10^{-12}\,\Omega$-cm$^2$, a value closer to the via resistance we calculate for fcc Ru. We see that even if a thin adhesion liner is required to grow Ru interconnects, we can expect dramatic benefits in vertical resistance compared to Cu interconnects.

\section{Conclusion}

Using \emph{ab initio} techniques, we benchmark the performance of fcc Ru for use in advanced interconnects. Although the hcp phase is the dominant one found in film measurements, the fcc phase is found to have a formation energy of 0.1 eV/atom more than the hcp phase, which may explain why it has been observed in confined structures. We find that the phonon-limited bulk resistivity of the fcc phase is three times lower than that of the hcp phase indicating that stabilizing fcc Ru can be beneficial for interconnect performance. When compared to Cu, however, we do see that increased grain boundary scattering results in fcc Ru having a higher bulk resistivity. We find, however, that the fcc Ru lines can offer lower line resistance compared to Cu lines below 21 nm line width due to the conductor volume lost to liner materials in Cu interconnects. In addition, we show that Ru may not require an adhesion liner by showing that the binding energy of fcc Ru to silicon dioxide is lower than that of many conventional adhesion liner materials. Even if a thin adhesion liner is required for BEOL integration, we calculate that the expected via resistance penalty can be up to eight times lower than that of typical liner and wetting layer stacks that are required for Cu integration. The combination of lower bulk resistivity compared to the hcp phase and dramatically lower via resistance compared to Cu suggest that the fcc phase of Ru could be a superior alternate conductor for future interconnect nodes.

\section*{Acknowledgments}

This  work  was performed by  the  Research  Alliance Teams  at various IBM Research and Development Facilities. The authors acknowledge D. C. Edelstein for insightful discussions.

%\section*{References}

% \bibliographystyle{aipnum4-2}
\bibliographystyle{apsrev4-1}
% \bibliographystyle{plain}
% \bibliography{../../../Latex/bibtex/Interconnects,BibMetals} 
\bibliography{Interconnects,BibMetals}

% \begin{thebibliography}{9}\label{sec:TeXbooks}%
% \bibitem{Note1}
% For help regarding the installation of this software and its use, please send email to \href{mailto:tex@aip.org}{tex@aip.org}.
% %
% \bibitem{Note2}
% Available with the  distribution, see \url{http://journals.aps.org/revtex/}.
% %
% \bibitem[Lamport(1996)]{LaTeXman} 
% L. Lamport, 
% \emph{\LaTeX\, a Document Preparation System} 
% (Addison-Wesley, Reading, MA, 1996).
% %
% \bibitem[Goossens(1994)]{Compan} 
% M. Goosens, F. Mittelbach, and A. Samarin, 
% \emph{The \LaTeX\ Companion} 
% (Addison-Wesley, Reading, MA, 1994).
% %
% \bibitem[Knuth(1986)]{TeXbook} 
% D. E. Knuth, 
% \emph{The \TeX book} 
% (Addison-Wesley, Reading, MA, 1986). 
% %
% \bibitem[Kopka(1995)]{Guide} 
% H. Kopka and P. Daly, 
% \emph{A Guide to \LaTeXe} 
% (Addison-Wesley, Reading, MA, 1995).
% %
% \bibitem[Goossens(1997)]{CompanG} 
% M. Goossens, S. Rahtz, and F. Mittelbach, 
% \emph{The \LaTeX\ Graphics Companion} 
% (Addison-Wesley, Reading, MA, 1997).
% %
% \bibitem[Rahtz(1999)]{CompanW} 
% S. Rahtz, M. Goossens \emph{et al.},
% \emph{The \LaTeX\ Web Companion} 
% (Addison-Wesley, Reading, MA, 1999).
% %
% \end{thebibliography}

\end{document}